Enhanced Fullerene Yield in Plasma-Aerosol Reactor at Cryogenic Boundary Temperature

Mikhail Jouravlev

**POSTECH, San 31, Hyojadong, Namgu, Pohang 790-784, South Korea,**
**(L. Ya Karpov Institute of Physical Chemistry, Vorontsovo pole 10, Moscow,**
**103064, Russia, in the past)**, e-mail: <u>jouravl@rambler.ru</u>

**Abstract.**

We demonstrate remarkably enhanced yield of $C_{60}$ fullerenes in an aerosol discharge chamber due to the additional presence of a strong spatial temperature gradient. The role of the temperature gradients in the increased yield of $C_{60}$ and fullerene-like structures is discussed. The reaction is not fully reversible and carbon soot matter is formed as a secondary product in the form of carbon aerosol particles. The increasing concentration of $C_{60}$ was easily recognized from the characteristic UV-spectra. The result of this paper will be useful for improvement of fullerene synthesis technology and for application to constructing new types of aerosol-plasma reactors.

**Keywords:** $C_{60}$, fullerenes, synthesis, kinetics, thermal treatment, temperature gradient, arc plasma, soot, carbon particles, combustion, graphite rod, plasma reactors, aerosols

**INTRODUCTION**

$C_{60}$ type fullerenes, owing to their unique molecular structure and low dimensionality, have excited much interest for their potential application to novel chemical technologies (1,2). However, the efficient production of macroscopic quantities of $C_{60}$ is still not possible. A main problem is the presence of impurities in the graphite rods used in the synthesis process. These impurities reduce the efficiency of fullerene production by means of participating as either free radicals or in ionic reactions during the combustion processes. In the synthesis of fullerenes by electrical discharge between graphite electrodes in an argon/helium environment, the presence of impurities is revealed by the detection of appreciable quantities of CO, $CO_2$ and $O_2$ (5,6).

The introduction of high spatial temperature gradients as a means of increasing the yield of fullerenes in plasma-arc discharge chambers has attracted significant attention. Here, we report our investigation of enhanced fullerene yield arising from an additional cryogenic boundary temperature gradient of the plasma chamber. The presented work is aimed to develop an additional scheme of $C_{60}$ fullerene synthesis for production of macroscopic quantities for novel material systems having useful physical and chemical properties (3,4). The yield is defined as the percentage of the net amount of $C_{60}$ (after evaporation/crystallization from solvents) with respect to the mass of crude materials i.e. the original graphitic soot (mass of slag after combustion of the rod) (3,4). The enhancement is defined as the relative yield by the ratio of the yield of $C_{60}$ production in the process of evaporation of the carbon road in the presence of high cryogenic boundary temperature gradient to the yield of $C_{60}$ production in the ordinary process of combustion.

**EXPERIMENTAL METHODS**

$C_{60}$ fullerenes were produced using the carbon-arc method (3,4) in a specially developed water-cooled aerosol-plasma chamber (4-6). To avoid oxygen contamination of the chamber, the graphite rod samples were kept under argon atmosphere prior to their evaporation. The evaporation occurred under helium atmosphere.

The aerosol-plasma chamber was cylindrical in shape with a height of 260 mm and a circular base with a diameter of 55 mm. The electrode junction was oriented perpendicular to the chamber axis. The graphite disk cathode of height 9 mm and circular base of diameter 40 mm was placed perpendicular to the chamber axis. The anode was a pure graphite rod similar to those used in the deceleration systems of nuclear reactors – its length and diameter were 210 mm and 6 mm respectively.

A plasma arc was initiated by electrical discharge under an electrode bias voltage of 25 V and arc current 115 A with the initial pressure of the argon/helium atmosphere at

0.2 atm. During evaporation, the graphite rod anode was moved by an automatic drive gear toward the cathode to maintain a constant arc current.

To introduce a strong spatial temperature gradient, the orifice of a liquid nitrogen Dewar flask at 78 K was connected to the top orifice of the aerosol-plasma chamber. The presence of the nitrogen Dewar flask reduces the partial pressure of volatile impurities in the main chamber – these impurities are condensed in the Dewar flask and cease to influence the fullerene synthesis. For example, the partial pressure of ionized gas in the chamber during the evaporation process in the presence of the strong temperature gradient rises from 0.2 atm to just 0.28 atm, compared to 0.38 atm when the liquid nitrogen Dewar flask is removed. Finally, the soot containing the fullerene molecules were extracted from the soot by a Soxhlet apparatus with toluene solution, and chromatographically purified by a column process (8).

## RESULTS AND DISCUSSION

It is known that $C_{60}$ can be destroyed by UV-radiation emitted by the discharge arc into the chamber. In our experimental setup, the UV arc radiation was shielded by the conjugation orifice connecting the Dewar flask with the main chamber. In addition, the Dewar flask provides the flow of gas-aerosol particles from the main chamber to the Dewar flask by the high temperature gradient of approximately 3000 K to 78 K.

About 70% of the soot produced in the evaporation process was found as sediment on the base of the main plasma-aerosol chamber. Thus the presence of the Dewar flask provides a means to extract combustion products from the main chamber that would normally inhibit the production of fullerenes. This reduces the overall synthesis time.

The soot formed during the evaporation process consisted of chaotically oriented, coagulated microparticles with sizes in the range 0.2 – 30 μm.

The average combustion time was 3 minutes. After synthesis, soot formed on the base of the main chamber by gravity sedimentation was collected.
Figure 1 shows the UV absorption spectra of the $C_{60}$ solution in toluene extracted from
the carbon soot produced during the arc synthesis under the high temperature gradient

(curve 1), and without the temperature gradient (curve 2). Curve 3 shows the UV spectra of the control solution of $C_{60}$ dissolved in toluene. Following the classical theory for the Lorenzian lines in the absorption spectra, the ratio of the amplitude of the curve 1 to the amplitude of the curve 2 provides the ratio of the numbers of $C_{60}$ molecules of one solution to the other. We estimate from the optical UV absorption spectra Fig.1, that the amount of $C_{60}$ in the solution extracted from the soot produced in the process of the combustion with the temperature gradient provided by the attached Dewar flask (amplitude of curve 1) was approximately 2.7 times more than the amount of $C_{60}$ in the solution extracted from the soot in the process without strong cryogenic boundary conditions (amplitude of curve 2).

The two additional peaks seen in curve Fig. 1 correspond to (A): $C_{70}$ at 378 nm and (B): $C_{60}$ at 404-408 nm. The small peak (A) of curve 1 corresponding to $C_{70}$ at 378 nm is obtained by analysis of the toluene extract from the soot mixture in full accordance with Ref.(7).

The analysis of the size distribution of aerosol particles in the soot indicates that coagulation of carbon particles occurs in the main chamber. We also have observed the relative yield of $C_{60}$ to be greater when the atmosphere in the chamber was helium, compared to argon.

**CONCLUSIONS**

The aim of this experiment was to determine the influence of temperature gradients on the yield of $C_{60}$ synthesized in an arc discharge chamber. It was shown that the presence of a strong temperature gradient produced by attaching a liquid nitrogen Dewar flask to a chamber orifice resulted in an increase of the $C_{60}$ fullerene relative yield by a factor of 2.7. We believe this increased relative yield is a result of the induced flow of impurities out of the chamber and into the flask where they are absorbed on the flask walls. One would expect that further decreasing the boundary temperature to 4.2 K, such as by using a liquid helium Dewar, would provide the possibility for further increase in the $C_{60}$ yield. In addition, the attached Dewar flask provides shielding of the produced fullerenes from UV arc radiation.

The method of high temperature gradients we have presented is a simple yet highly beneficial modification to the well known synthesis methods of fullerene production (1-4). This research has provided a new method for the high yield synthesis of fullerenes. A possible extension of this method is to change the simple orifice and connection tubes of the aerosol chamber by the Laval's nozzle for the creation of both temperatures gradients and gradients of the pressure.

## ACKNOLEDGEMENT

This work was carried out at Aerosol Technology Lab. of Karpov Institute of Physical Chemistry, Moscow, where the author was supported by the Karpov fellowship award. The author acknowledges N.N. Belov and Daniel R. Mason for the useful discussions.

## REFERENCE

1. Kroto, H. W.,  Heath, J. R., O'Brien, S. C., Curl, R. F., and Smalley, R. E. (1985) $C_{60}$: Buckminsterfullerene. *Nature* **318**, 162 - 163

2. Kratschmer,W., Lamb, L.D., Fostiropoulos, K., and Huffman, D.R. (1990) Solid $C_{60}$: a new form of carbon.  *Nature* **347**, 354 - 358

3. Haufler, R.E., Concencao, J., Chibante, P. et al. (1990) Efficient production of $C_{60}$ (Buckminsterfullerene), $C_{60}H_{36}$, and the solvated buckide ion. *J. Phys. Chem.* 94, 8634 - 8636.

4. Parker, D.H., Wurz, P., Chatterjee, K. et al. (1991) High-yield synthesis, separation, and mass-spectrometric characterization of fullerenes $C_{60}$ to $C_{266}$. *J. Am. Chem. Soc.* 113. 7499-7503.

5. Belov, N.N., Simanchev, S.K., and Tokarevskikh, A.V. (1997) Alteration of electrodes surface during fullerene synthesis by arc. *J. Aerosol. Sci.*, 28, Suppl.1: S533-S534.

6. Belov, N. N., Simanchev, S. K. and Tokarevskikh, A. V. (1997) Structure and Arrangement of Carbon Microparticles of Electrodes during Fullerene Synthesis in Inert Gases (Ar, He). *Fullerenes, Nanotubes and Carbon Nanostructures,* 5:7, 1449-1460

7. Ajie, H. et all. (1990) Characterization of the soluble all-carbon molecules $C_{60}$ and $C_{70}$. *J. Phys. Chem.* 94, 8630-8633.

8. Hirsch, A., Brettreich, M., (2005) Fullerenes: Chemistry and Reactions. Wiley-VCH Verlag GmbH&Co. KGaA. Weinheim. p.26

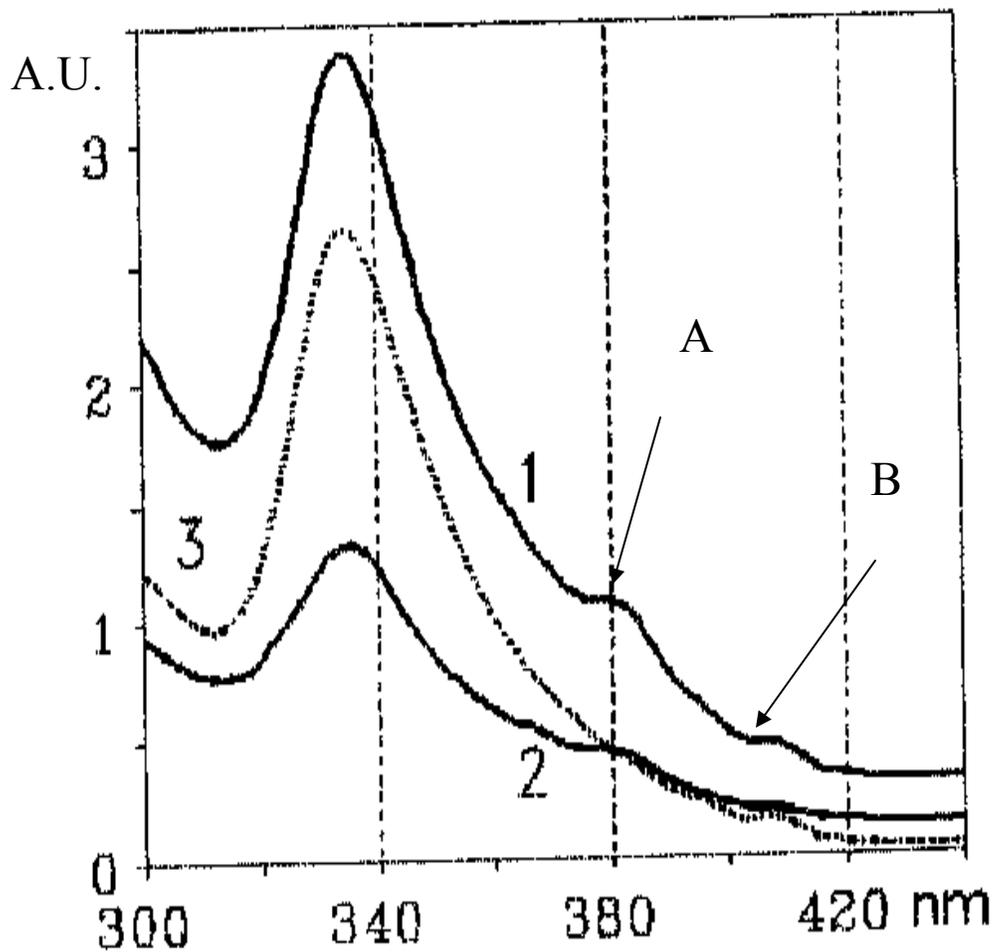

Fig.1 The UV absorption spectra (optical intensity in arbitrary units versus wavelength) of extracted $C_{60}$ containing soot dissolved in toluene produced for the case of high gradient temperature field in the plasma-aerosol chamber (curve 1), no temperature gradient (curve 2), and the control solution of $C_{60}$ in toluene (curve 3). Peak (A): $C_{70}$ at 378 nm, and peak (B): $C_{60}$ at 404-408 nm.